\documentclass[aps,prl,twocolumn,showpacs,nofootinbib,floatfix]{revtex4-1} 
\usepackage{graphicx}  
\usepackage{amssymb}
\usepackage{epstopdf}
\usepackage{color}
\usepackage{array}
 
\begin{document}

\title{\bf Transition probabilities in the X(5) candidate $^{122}$Ba}
\author{ P. G. Bizzeti$^{1,2}$, A. M. Bizzeti--Sona$^1$, D. Tonev$^{3,4}$, 
A. Giannatiempo$^{1,2}$,  C.A. Ur$^{5\dag}$, A. Dewald$^6$, 
B.~Melon$^{1,2}$, C.Michelagnoli$^{5,9}$,  P. Petkov$^4$, D. Bazzacco$^5$, 
A. Costin$^{10}$, G.~de~Angelis$^3$, 
F.~Della~Vedova$^{3}$, M.~Fantuzi$^{7,8}$, E.~Farnea$^{5}$, C.~Fransen$^6$,
A. Gadea$^{3,11}$, 
S.~Lenzi$^{5,9}$, S.~Lunardi$^{5,9}$, N.~Marginean$^{3\dag\dag}$,
 R.~Marginean$^{5\dag\dag}$,  
R. Menegazzo$^{5}$, D. Mengoni$^{5,9\dag\dag\dag}$, 
O.  M\"oller$^{10}$,  A.~Nannini$^2$, D.R. Napoli$^3$, M. Nespolo$^{5,9}$, 
P.~Pavan$^{5,9}$,  A. Perego$^{1,2}$, 
 C.M. Petrache$^{7,8\dag\dag\dag\dag}$,  N. Pietralla$^{10}$, C. Rossi Alvarez$^5$,
P. Sona$^{1,2}$
\vspace{0.3cm}}

\affiliation{\it
\bigskip
$^1$Dip. di Fisica e Astronomia, Universit\`a di Firenze;
$^2$ INFN, Sezione di Firenze;
$^3$INFN, Laboratori Nazionali di Legnaro;
$^4$Institute~for~Nuclear Research and Nuclear Energy, BAS, Sofia; 
$^5$INFN, Sezione di Padova;
$^6$Institut f\"ur Kernphysik der Universit\"at zu K\"oln; 
$^{7}$Dip. di Fisica, Universit\`a di Camerino;  
$^{8}$INFN, Sezione di Perugia;
$^{9}$Dip. di Fisica, Universit\`a di Padova; 
$^{10}$Institut f\"ur Kernphysik, TU Darmstadt.
$^{11}$IFIC, CSIC -- University of Valencia,~Spain.
$^\dag$On leave from IFIN-HH Bucharest.
$^{\dag\dag}$ Now at IFIN-HH Bucharest.
$^{\dag\dag\dag}$ Now at University of the West of Scotland, Paisley, UK.
$^{\dag\dag\dag\dag}$Now at IPN Orsay, IN2P3-CNRS and Universit\'e 
Paris-Sud, France.
 }

\begin{abstract}
To investigate the possible X(5) character of  $^{122}$Ba, suggested by the
ground state band energy pattern,  the lifetimes of the lowest yrast 
states of $^{122}$Ba  have been measured, {\it via} the Recoil Distance
Doppler-Shift method. The relevant levels have been populated by using 
the $^{108}$Cd($^{16}$O,2n)$^{122}$Ba  and the
 $^{112}$Sn($^{13}$C,3n)$^{122}$Ba reactions. 
The B(E2) values deduced in the present work are compared to the predictions
 of the X(5) model and to  calculations performed in the framework of 
 the IBA-1 and 
IBA-2 models.
\end{abstract}


\maketitle

\section{Introduction} 

The  X(5) model, introduced  by  Iachello ~\cite{iac02} in 2001,  describes
nuclei close to the critical point of the phase transition between spherical
and  axial prolate shape. Many theoretical and experimental efforts have 
been already dedicated to the study of the 
X(5) critical point phenomenon
\cite{Casten-01,Kruecken-02,Bizzeti-02,Hutter-03,Fransen-04,
Caterina,Tonev-04,Caprio-04,McCutchan,Pietralla-04,Dewald-05,Balabanski-06,
 Bonatsos-07,Mertz-08}.
Recent experiments provided evidence that, while several nuclei have a level
scheme in agreement with the X(5) model, only a few  among them have $E2$ 
transition strengths which
also agree with the model predictions (see, {\it e.g.}~\cite{clark03}).

\begin{figure}
\includegraphics[width=9.cm,clip,bb=54 421 574 787] 
{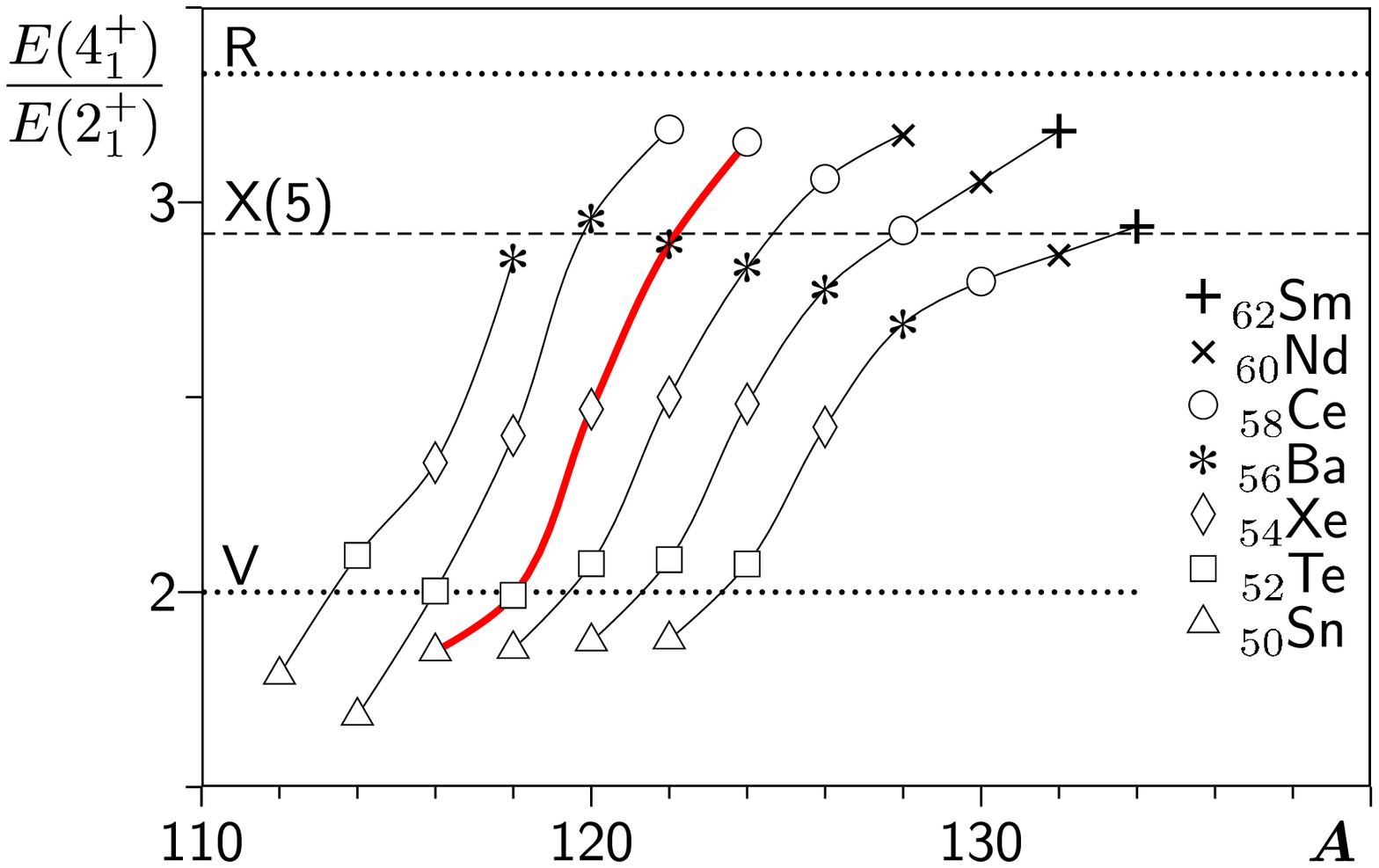}
\caption{\label{F:1}(Color online) Evolution of the ratio 
$E(4^+_1)/E(2^+_1)$ as a function of the proton number in the isotone 
chains with $N=62,64,66,68,70,72$
 (from left to right). The data are reported 
versus $A=N+Z$, to avoid  overlapping of the different curves. A thicker 
line outlines data pertaining to the $N=66$
isotone chain. Horizontal lines show the theoretical values predicted by
harmonic vibrator (V),  X(5) model, and  rigid rotator (R).}
\end{figure}

The light Ba isotopes are known to belong to a transitional region  between
spherical and axially deformed nuclei, as
shown in Fig.~\ref{F:1}, where the ratios $E(4^+_1)/E(2^+_1)$  for
$N=62-70$ isotones are reported.

Among them, $^{122}_{\phantom{1}56}$Ba$^{\phantom{66}}_{66}$ has been
 proposed by Fransen {\it {\it et al.}}~\cite{Fransen-04} as a rather
 good X(5) candidate on the basis of the  agreement observed between
 experimental  and predicted level  energies. 

 The main purpose of this work is to test the possible X(5) character
 of this nucleus by measuring the lifetimes in the ground state (g.s.)
band in order to deduce  the $E2$  strengths of the transitions de-exciting
 its levels and,  possibly, to identify the excited $\beta$ band
 (the $s=2$ band according to X(5)  terminology).    

 Most of the spectroscopic information on $^{122}$Ba available in
 the literature has been obtained from fusion--evaporation
reactions. In an old work  the ground-state band was observed 
 up to $J=12$ ~\cite{conrad}, while in  two more recent
works~\cite{petrache,Fransen-04}  the study has been  extended to high spin
 states and several bands have been identified. 

As to the  electromagnetic properties, only the lifetime of the  $2_1^+$ 
state has been measured (428$\pm$39 ps) from delayed $\beta$-$\gamma$ 
coincidences  in the decay of $^{122}$La \cite{Morikawa-92}.

To gain additional spectroscopic information on  $^{122}$Ba three series
of measurements have been performed. The first two experiments have been 
carried out at the XTU Tandem accelerator
 of Laboratori Nazionali di Legnaro (LNL)  with the GASP array~\cite{GASP},
  by means of the $^{108}$Cd($^{16}$O,2n)$^{122}$Ba reaction.
  A thick  target was used in the first case, in order to obtain  high 
 statistics of double and triple $\gamma$ coincidences, with the main  aim of
 identifying the $0^+$ band head of the $s=2$ band. In the second one,
 the lifetimes of  the excited levels were measured by  the Recoil Distance
Doppler--Shift (RDDS) Method, using the Plunger device of the Cologne group.
The third measurement, performed at the  Tandem accelerator of the Institut
 f\"ur Kernphysik, Cologne,
using the  $^{112}$Sn($^{13}$C,3n)$^{122}$Ba reaction, was  intended to
 obtain an independent
check for the lifetime of the first  excited level, which provides
 the reference value for the normalization of the $B(E2)$ strengths.

Some results of a preliminary evaluation of data have been presented at the
Conference CGS13  (Cologne 2008)~\cite{colonia}. Here, we report the final
 analysis and results, which supersede the previous ones.

\section{Experiment} 

\subsection{Thick--target measurement} 

The level scheme of $^{122}$Ba has been investigated {\it via} the 
$^{108}$Cd($^{16}$O,2n)$^{122}$Ba reaction at  62 and 64 MeV beam energy. 
The target consisted of a 1~mg/cm$^2$, 69\% enriched $^{108}$Cd foil,
followed by a 10~ mg/cm$^2$ $^{208}$Pb backing. The GASP array was mounted
 in  its Configuration II, {\it i.e.}, with the Compton suppressed
 Ge  counters at an average distance of 22 cm from the  target. 
Triple and higher--fold coincidence events have been recorded in  55~h 
 of measurement. From the
raw data, we have constructed three- and two-dimensional coincidence
  matrices, containing approximately 
$1.9\times 10^7 ~  E_\gamma -E_\gamma-E_\gamma $  coincidences and 
 $1.5\times 10^8$   $E_\gamma -E_\gamma$ coincidences.
These matrices have been used for the search of weak $\gamma$ transitions
 in  coincidence with known transitions in $^{122}$Ba.

\begin{figure}
\includegraphics[width=8.5cm,clip]
{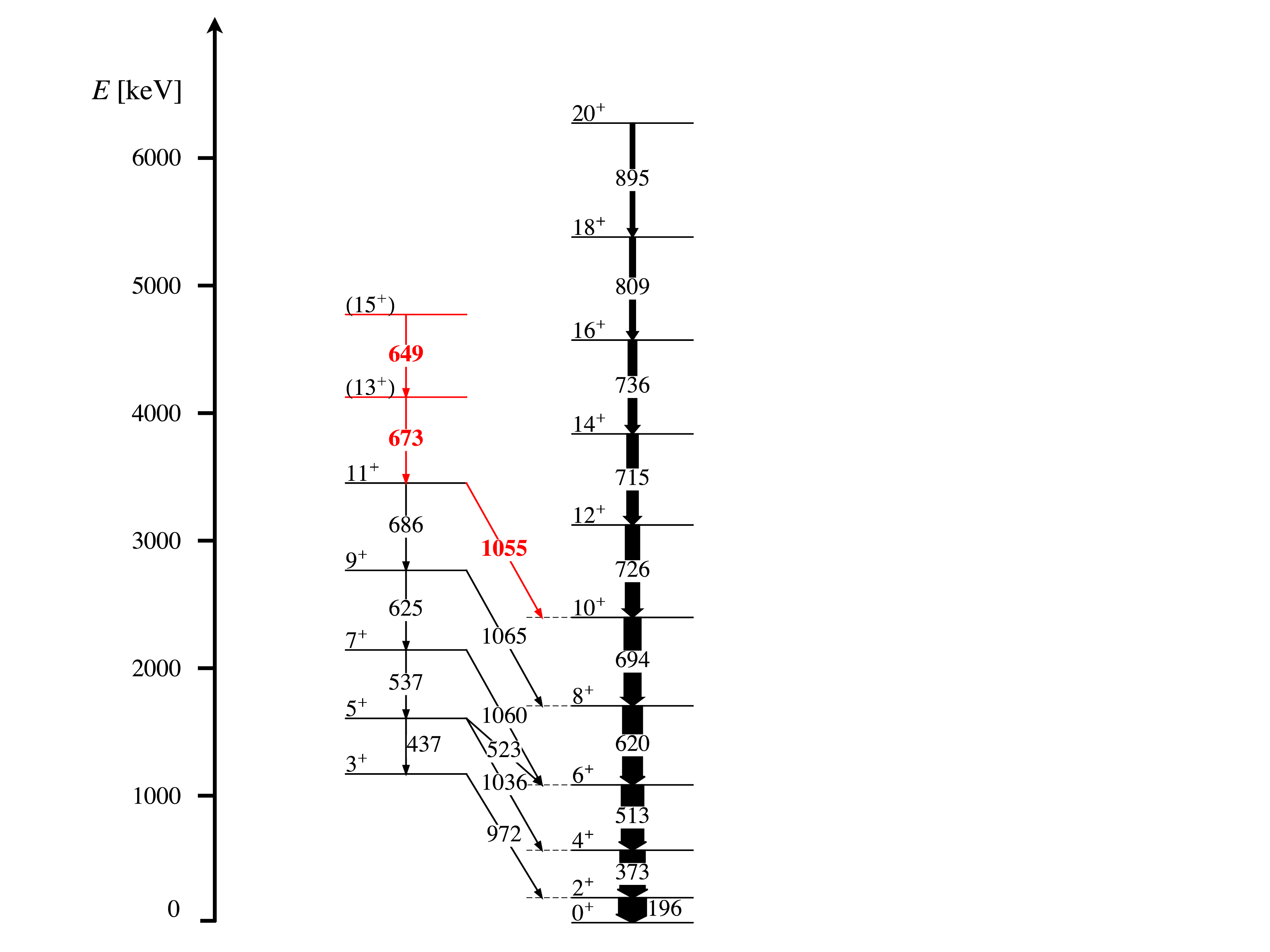}
\caption{\label{F:2}(color online)
Partial level scheme of $^{122}$Ba  relevant for the present analysis.
 The two  levels on the top of the band based on the $3^{+}_{1}$ state
 and the $11^{+}_{1} \to 10^{+}_{1}$ transition have been identified
 in the present work.}
\end{figure}

Most of the transitions belonging to the bands reported in the previous works
  \cite{petrache,Fransen-04} have been identified. A revised version of
 the excitation energy pattern  of   $^{122}$Ba will be presented in
 a forthcoming paper which will also include  the lifetime measurements of
 the  yrast band  levels of J $\ge$ 10  \cite{mic2010}. Here,  we restrict
 the analysis to the known positive parity bands. 

The yrast $\gamma$ cascade, reported  in  \cite{petrache} (see Fig. 2)
  has been  observed at least up to the $20^+_1 \to 18^+_1$ transition.
As to the band based on the $3^+_1$ state at 1168 keV, reported by Fransen
 {\it {\it et al.}} \cite{Fransen-04},  it has been extended by two more
 transitions.
   They show  the typical angular distribution of stretched E2 transitions
 and therefore the suggested spin for the corresponding levels are 13
 and 15. In addition, a new inter-band transition connecting 
 the $11^{+ }_1$ and  $10^{+}_1$ states has been found.

Fransen {\it et al.}~\cite{Fransen-04} have also observed two $\gamma$
 transitions, of 587 keV  and 422 keV, in 
coincidence with one another and with a 196 keV transition which could
 originate from the first excited $J=2^+$ state of $^{122}$Ba.
In the present measurement, it has been possible to  extend  the 196,
 422, 587 keV cascade to higher lying levels, decaying with $\gamma$
 transitions
of 722 keV, 838 keV and 880 keV    (in order of decreasing intensity).
 We have identified  the entire cascade as a part of a known band
  of $^{121}$Xe, based  on the $7/2^-$ level at 196 keV~\cite{xe121}. 
Indeed, the $^{121}$Xe nucleus  is populated in the n2p exit channel of 
the used reaction.

Moreover, we have observed two transitions, of  199 keV and 514 keV
(belonging to the band (4) reported by Petrache 
{\it et al.}~\cite{petrache}),
 which are close in energy  to those of the first and third transition of
 the ground-state band (196 keV and  513  keV, 
respectively), and are in coincidence with the transitions de-exciting 
 the $8^+_1$ and the $6^+_1$ levels. In the following, for the sake of
 simplicity, they will be mentioned as 199 keV-band(4) and 514 keV-band(4)
 transitions.

 In view of the analysis of the data obtained with the  RDDS method
 (described in the following section), for which an overlap of the
 514 keV-band(4)  and  513 keV transitions could be somewhat problematic,
we have determined the relative intensity of the two transitions. 
They are in coincidence with the $8^+_1\to 6^+_1$, 620 keV and  the 
$4^+_1\to 2^+_1$, 373 keV transitions.  Through an accurate analysis, where
 a reference transition has been exploited,  it has been found that
  the intensity ratio of the 514 keV-band(4) to the 513 keV line,
 in coincidence with both the 620 keV and 373 keV transitions,
 is $(1.6\pm 0.5)\ 10^{-2}$, while it is larger ($5.0\pm 1.4\ 10^{-2}$)
 when in coincidence with only the 373 keV transition, due to the fact that
 several transitions from band (4) of  \cite{petrache} directly feed
the $6^+_{1}$ or the $4^+_{1}$ level.  Since all the transitions of 
interest are stretched E2, we expect that their intensity ratio be almost 
independent of the detector angle.

We have also checked that a weaker $14^-_2\to 13^-_2$, 371 keV transition
 in band (3) of \cite{petrache} has practically no effect on our results.

As to the $0^+_{2}$ state (the band head of the $\beta$ band),
  the  systematics of heavier barium isotopes
(see, {\it e.g.},~\cite{Fransen-04,Caterina}) would suggest  an excitation 
 energy  of about 1.1 MeV, which is very close to the X(5) prediction.
Particular care has been devoted to the search, in the coincidence data, 
of possible $\gamma$ rays de-exciting states belonging to the $\beta$ band.
In particular,   the energy spectrum gated on the $2_1^+ \rightarrow 0_1^+$
 transition  has been thoroughly analyzed, but no indication for such 
$\gamma$ ray has been found. At the moment, the  $\beta$ band of $^{122}$Ba
  is still   to be identified.

\subsection{Lifetime measurements}

In a separate experiment, performed at LNL, the lifetimes of  the excited
 states  of the  g.s. band in $^{122}$Ba  have been measured by means of
the RDDS  method, with the Plunger device of the Cologne  group   installed
 in the GASP array  (in Configuration 2). Counter rings 1, 2, 6, and 7 (at 
 $34.6^\circ,\ 59.4^\circ,\ 120.6^\circ$ and $145.4^\circ$ to the beam
 direction, respectively) provided sufficient  separation of the
 Doppler-shifted  part and the unshifted part for most of the $\gamma$-ray
 lines.
Excited states in $^{122}$Ba were populated {\it via} the reaction
 $^{108}$Cd($^{16}$O, 2n)$^{122}$Ba. The target consisted of 0.5  mg/cm$^2$
$^{108}$Cd   on a 2.3 mg/cm$^2$ Ta foil (facing the beam).  A 6.4 mg/cm$^2$
 gold foil was used to stop the recoils.
The beam energy was 69 MeV and was reduced to about 65 MeV at the entrance
 in the $^{108}$Cd target. Measurements have been performed at 17  
 target-to-stopper distances, in the range from point of electric contact
between target and stopper and $1389~\mu$m.
 Since the average  velocity of the recoiling nuclei  was 
$v \approx 10^{-2} c$, the corresponding values for  the time of flight 
were in the range $\approx
 0.5$~ps -- $ 463$~ps. 

Typical examples of the evolution of the line shape with the 
target-to-stopper distance are shown in  Fig. 3.

\begin{figure}
\includegraphics[width=7.5cm]  
{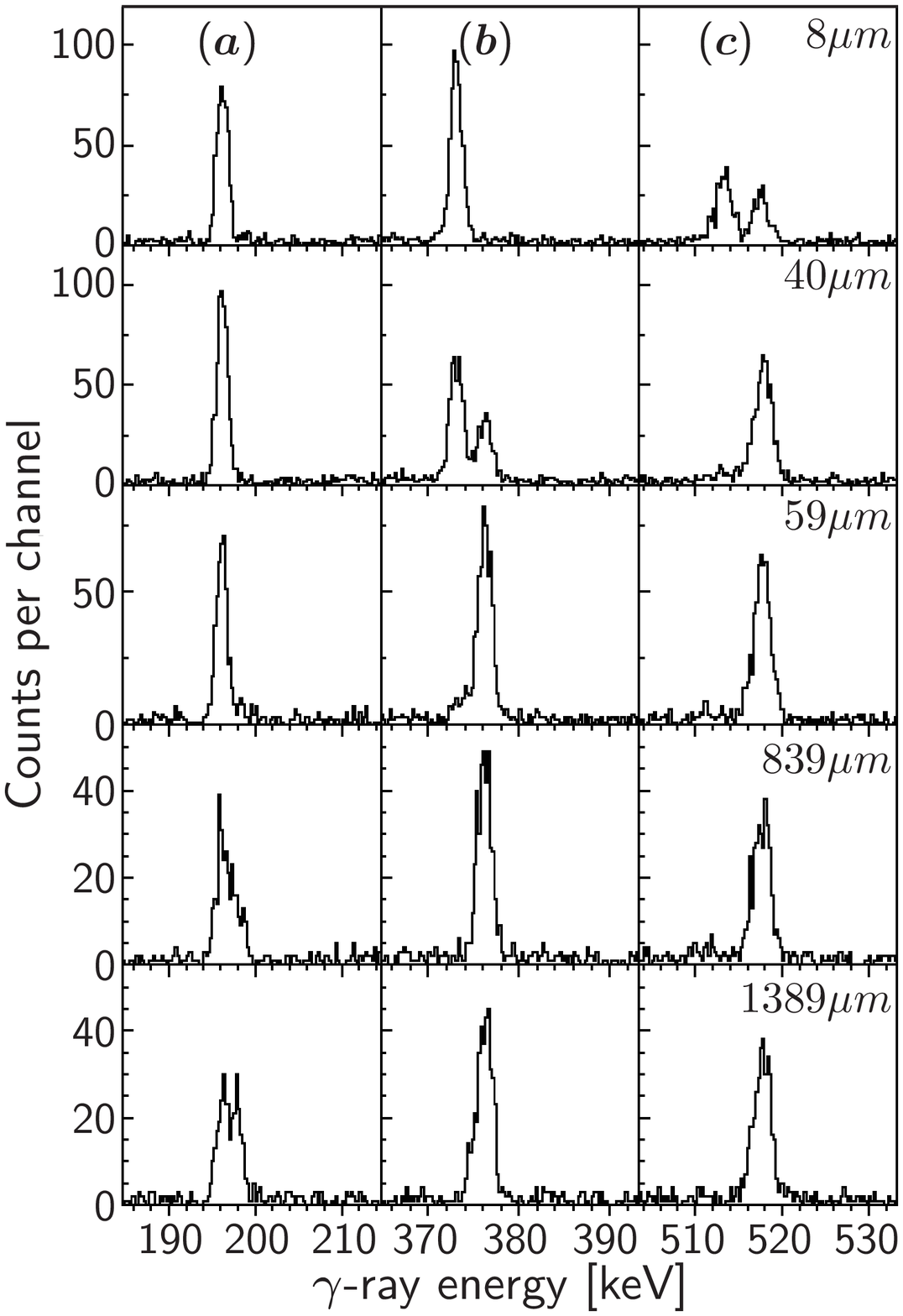}
\caption{\label{F:3} Evolution of the line shape of  the 
$2^+_1\rightarrow 0^+_1$ of 196 keV ($a$), $4^+_1\rightarrow 2^+_1$ of 373 
keV ($b$), and $6^+_1\rightarrow 4^+_1$ of 513 keV ($c$)  $\gamma$ 
transitions at different target-to-stopper distances, in the experiment 
performed at LNL. These spectra have been taken  at the most forward ring 
(34.6$^\circ$ with respect to the beam), in coincidence with the 
Doppler-shifted part of the directly feeding  transition. The size 
of the small 511 keV peak is visible in  the lowest part of the 
right-hand panel.}
\end{figure}
As to the $2^+_1\rightarrow 0^+_1$ transition, it is seen that, even at 
the largest distances, the intensity of the unshifted line is still 
large and that of the Doppler  shifted line is still far from saturation. 
To extend the data to  larger distances, thus improving the reliability 
of the result, a new measurement has been performed {\it via} 
the $^{112}$Sn($^{13}$C,3n)$^{122}$Ba reaction, at the Tandem  Accelerator 
of the IKP of the  University of Cologne.
 The target consisted of a 0.3~mg/cm$^2$ $^{112}$Sn layer, evaporated on a 
1.7~mg/cm$^3$ Ta backing (facing the beam). 
A 4.0 mg/cm$^2$ gold foil was used to stop the recoiling nuclei.
The beam energy was 61 MeV (corresponding to 59 MeV at the entrance in the 
Sn target). In  this case, the average recoil velocity was 
$v \approx 8\times 10^{-3} c$.
A total of 11 Ge counters were mounted in two  rings, 5 at 143 degrees and 
6 at 29 degrees, while a further counter was placed at 0 degrees 
with respect to the beam direction. 
The average distance of the counters from the target was 11 cm for the 
backward ring and 20 cm for the forward one. 
A Plunger device \cite{Dewald-92}, like in the case of the previous 
measurement, was  used. Measurements have been performed at 8 
target-to-stopper distances ranging from 10~$\mu$m to 3000~$\mu$m, 
so that the significant part of the decay curve of the $2_1^+$state 
was almost  completely included.
The quality of the data obtained   from  this experiment is illustrated 
in Fig. 4.
 In spite of the relatively small average recoil velocity  the evolution 
of the unshifted and shifted peaks with increasing 
target-to-stopper distance is clearly    visible. 

For the data analysis, the Differential Decay Curve method 
(DDCM)~\cite{Dewald-89,Boehm-93}  (described in the next Section) has been 
used. When necessary, the finite stopping time of the ions in the stopper
material (Fig.~\ref{F:6a}) has been taken into account.

Moreover, a new method, the Integral Decay Curve Method (IDCM) \cite{IDCM} 
has been exploited in the analysis of the data collected at LNL, concerning
the lifetime of the 2$_{1}^{+}$. Such a method should be particularly 
advantageous when  the most significant points  for lifetime evaluation 
are close to the upper border of the explored range of distances, as in the
case considered
\footnote{\label{fn:1}This method, in fact, should be free of uncertainties
related to the evaluation of derivatives  at the border of the region. 
In addition, it 
takes automatically into account possible effects due to the angular and  
velocity spread of the moving nuclei.}.

\section{\label{Dataar}Data analysis and results}

In the RDDS measurements, the lifetime determination is based on the 
precise knowledge of the areas of the shifted (S) peak, corresponding to 
the emission of a $\gamma$-ray depopulating the investigated level while the
recoil nucleus is  flying  in vacuum or is slowing-down in the target and 
stopper, and  of the unshifted (U) peak, corresponding to an emission when 
the recoil is at rest in the stopper.
In such experiments, the evolution of the intensity splitting between these
two components, as a function of the target-to-stopper distance, is 
sensitive to the lifetime $\tau$ of the depopulating level. 
Such a splitting is shown in Figs. 6-9. 
 
For the lifetime determination, we used the differential decay curve method,
proposed in~\cite{Dewald-89,Boehm-93}. 
According to this method, at each target-to-stopper distance $x$, the 
lifetime $\tau(x)$ of the level of interest is deduced  from 
quantities obtained directly from the measured data.  The method can be 
applied for data analysis of singles as well as coincidence measurements. 
A crucial factor for the lifetime determination in RDDS measurements is the
precise knowledge of the feeding of the investigated state. In the present
analysis lifetimes are always determined using a procedure where a gate is 
set on the shifted component of a feeding transition. In this way the 
problem related to an unobserved feeding  is completely solved. 
Moreover, by gating only on the shifted component of  a feeding transition 
the effect of nuclear de-orientation cancels out  completely and does not 
influence the results of the  lifetime  analysis~\cite{Petkov-94}.
Therefore, special attention has been payed to put a gate only on the 
shifted component of the feeding transition.
When this component was not well separated from the unshifted one, only the
part  corresponding to the larger values
of Doppler shift has been considered. This  means to select a fraction of 
recoil nuclei with an exactly fixed range of the velocities.
This fact has  been taken into account in the data analysis  following the 
procedure  described in \cite{Boehm-93}.
In the choice of the gate, the same range of recoil velocities has been used
for the different rings.

\begin{figure}
\includegraphics[width=8.5cm,clip,bb=137 287 470 760]
{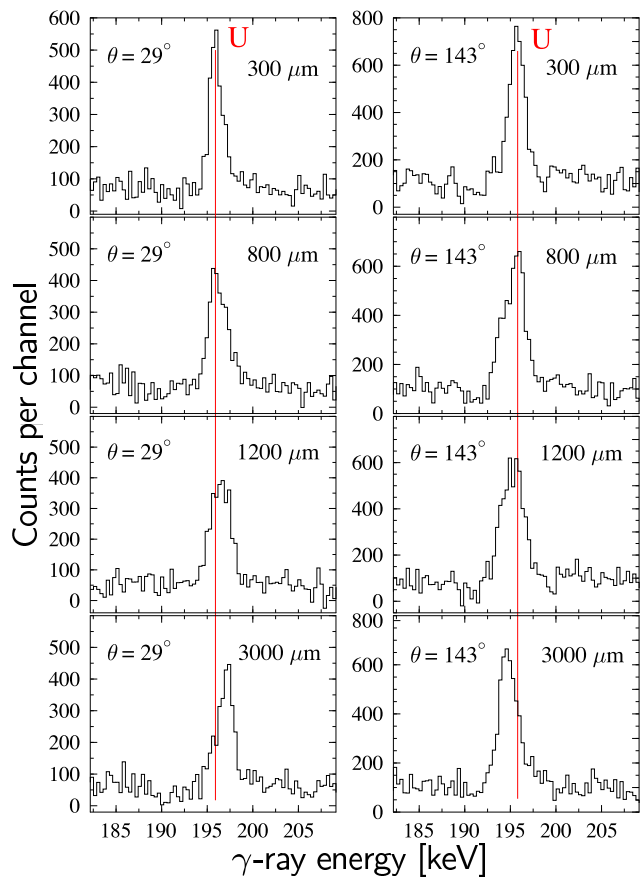}
\caption{\label{spectra}(Color online) 
Gated $\gamma$-ray spectra of the 2$_{1}^{+}$ $\rightarrow$
0$_{1}^{+}$ transition in the experiment performed in Cologne. On the 
left-hand side, the evolution with the distance of the shifted and unshifted
peaks of the 2$_{1}^{+}$ $\rightarrow$ 0$_{1}^{+}$ transition registered by
detectors positioned at an angle of 29$^{\circ}$ 
with respect to the beam is shown.
On the right-hand side is presented the same evolution registered by 
detectors positioned at a backward angle of 143$^{\circ}$.  The position 
of the unshifted (U) peak is indicated with a vertical line.}
\end{figure}

In the case of coincidence measurements, as in our experiments, the lifetime
 can be obtained according to the equation:
\begin{equation}\label{normal-DDCM-tau}
\tau(x) = \frac{\{B_{s},A_{u}\}}{\frac{\rm d}{{\rm d}x} \{B_{s},A_{s}\}} 
\frac{1}{<v>}
\end{equation}

Here, $<v>$ is the mean velocity of the recoils, and the quantities in 
braces are the numbers of experimental coincident events detected for the
shifted { ($A_s$) and unshifted ($A_u$) part of the  transition ($A$) 
involved in the analysis, in coincidence with the shifted part of a directly
feeding transition ($B_s$).   The derivative in the denominator of 
Eq.~\ref{normal-DDCM-tau} is obtained by fitting a quadratic spline to the 
intensity of the gated S component.

The precision of the evaluation of the areas of the U-  and S-peaks  sets 
limitations on the precision of the investigated  lifetimes. The correct 
determination of these areas, however, depends on several factors, discussed
extensively in  \cite{Petkov-99-A}. Here we only mention the main points 
which are relevant to the present measurement. 
In our experiment the beam energy, chosen to ensure a maximal cross-section
for the nucleus of interest, leads to
recoil velocities v/c around 1\%. In such a case the shifted and  unshifted
peaks for low energy transitions are not well separated and we need to 
know exactly the lineshapes of the two peaks in order to determine 
their areas.
Whereas the shape of the U-peak is described by the response function of the
detector, the shape of the S-peak depends also on 
all those factors (target thickness, stopping powers, reaction used, beam 
energy) which determine the velocity distribution of
the recoiling nuclei. Also a change in the target-to-stopper distance leads
to a modification in the line shape of the shifted peak: the faster
recoils reach the stopper earlier and have a higher probability to emit 
gamma-rays at rest than the slower ones, therefore faster recoils contribute
more to the unshifted peak. 

For the shorter lifetimes, ({\it i.e.}~$< 2$~ps)} it is important to take 
into account the  finite slowing-down time of the recoils in the stopper. 
During 
the slowing-down, the emission of de-exciting $\gamma$-rays  leads to the 
appearance of a continuous DSA (Doppler Shift Attenuated) component in the 
composition of the line-shape  which can be described by the procedure 
published in ~\cite{Petkov-99-A}.
 As reported  in some cases of the $A\approx 130$  mass 
region~\cite{Klemme-99,Petkov-00}, taking  into account this effect in the 
analysis 
has led to a correction of the lifetime values  (up to 70 \% for very short
lifetimes). The importance of the DSA effect in the present measurements is
evident in Fig.~\ref{analysis_620}, which shows data concerning the lifetime
determination  for the 8$^{+}_{1}$ state in $^{122}$Ba. For the 8$^{+}_{1}$
(10$^{+}_{1}$) state, the deduced   lifetime (limit)  turns out to be about
15$\%$ longer once the DSA effect has been taken  into account.

\begin{figure}[t]
\includegraphics[width=8cm]
{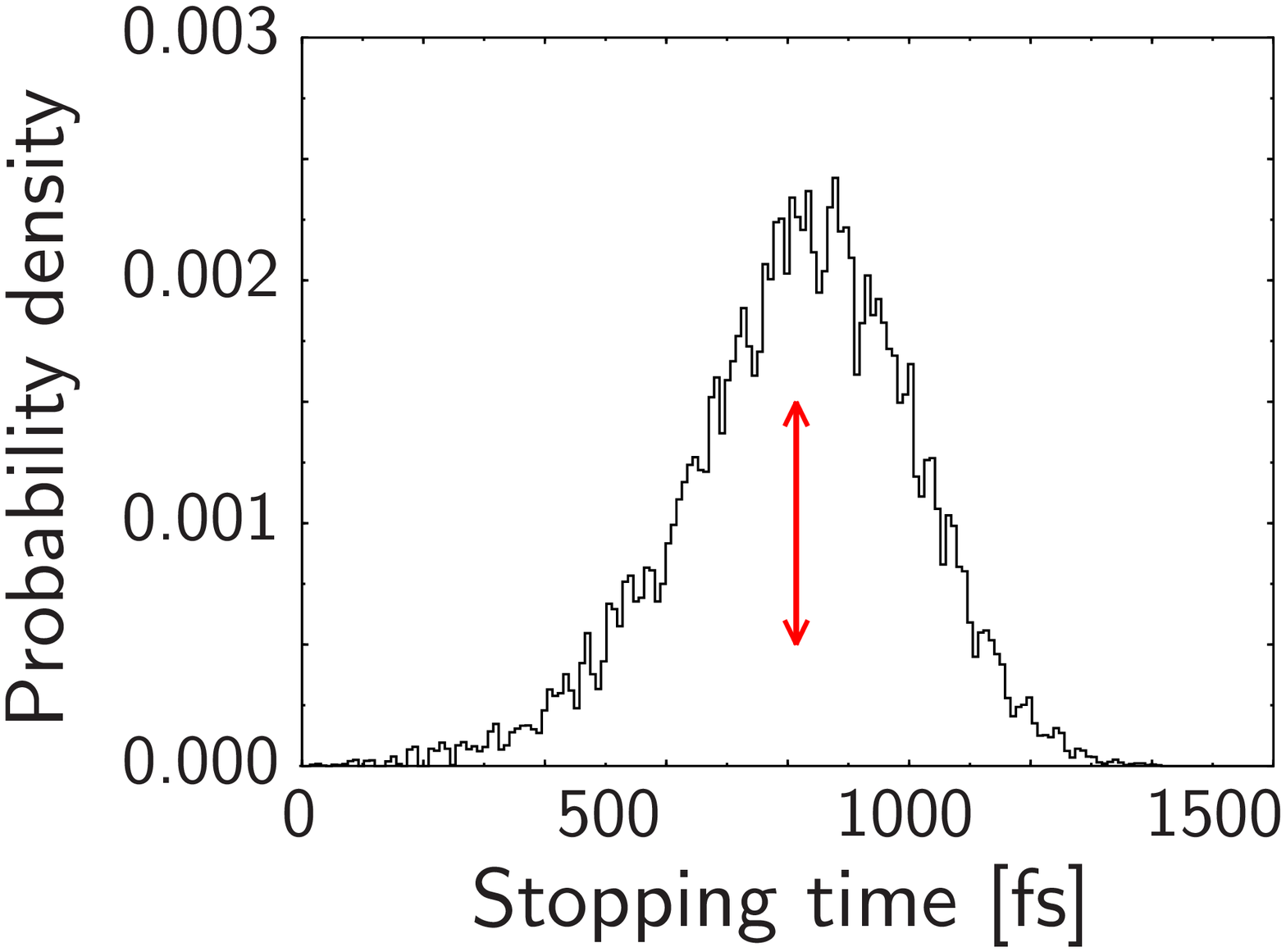}
\caption{\label{F:6a} (color online)
Distribution of the stopping time of the $^{122}$Ba recoils according to the
predictions of the MonteCarlo simulation for 10000 histories. The velocity 
distribution at the entrance of the gold stopper is calculated for a $^{108}
$Cd target thickness of 0.5 mg/cm$^2$ (on a 2.3 mg/cm$^2$ Ta facing the 
beam) and a beam energy of 69 MeV at the entrance of the target. The 
position of the mean value of the stopping time (814 fs) is indicated  a
vertical arrow. }
\end{figure}

The precise knowledge of the time evolution of the velocity distribution of
the recoils is critical for the description of the line shapes. The best way
 to obtain this  information is by performing a full  Monte Carlo (MC) 
simulation of the creation of the recoils, slowing down in the target, free
flight in vacuum, and slowing down in the stopper. At a second stage of the
MC simulation, the velocity  histories are randomized with respect to the 
two detectors between which the $\gamma$-$\gamma$ coincidences are recorded.
Numerical simulations reveal that a summation over 10000 MC-histories is 
sufficient for a stabilization of the line shapes. 
We performed a Monte-Carlo simulation  of the slowing-down histories of the
recoils using a modified~\cite{Petkov-98,Petkov-99-A} version of the program
DESASTOP~\cite{Winter-83} which allows for a numerical treatment of the 
electron stopping powers at  ion energies relatively higher  than those of
the original version. On this basis, a procedure for the analysis of 
the measured data was developed
\cite{Petkov-98,Petkov-99-A}. In order to fix the stopping power parameters
of the target we  considered  the spectra at large distances, where only 
shifted peaks are present for shorter lifetimes. Their line-shapes  are 
satisfactorily reproduced by using the simulated velocity distribution. 
Hence one can conclude that the stopping power of the target material are 
correctly taken into account. The effect of the momentum carried out by 
evaporated neutrons from the compound nucleus is also taken into account in
the MC procedure. Concerning the slowing-down in the stopper, for the 
experiment performed at LNL, the mean time interval needed by the recoils to
come to rest is predicted to be about 0.8 ps while the whole process is 
fully completed  within about 1.4 ps (see Fig.~\ref{F:6a}). 
The time step used for the digitalization of the velocity histories was 
$\Delta t$ = 8.87 fs.

 To illustrate the application of the procedure followed in the analysis, 
we show in Figs. 6-9 examples concerning the lifetime determination for the
196 keV, 373 keV, 513 keV, and 620 keV  transitions  which depopulate the 
$J^{\pi} = 2_{1}^{+},  4_{1}^{+}$,  $6_1^+$, and $8_{1}^{+}$ levels, 
respectively. On the left-hand side of the figures, the line shapes, recorded
at the  indicated distances, are displayed together with the fits. 
We would like to state that  the lifetime value reported in each  figure is
obtained from the data collected in the ring mentioned in the caption, 
whereas the corresponding one given in Table \ref{table_values} is an 
average value deduced from the data coming from all the relevant rings.

\begin{figure}
\includegraphics[bb=5 30 585 800,clip,width=8.64cm]
{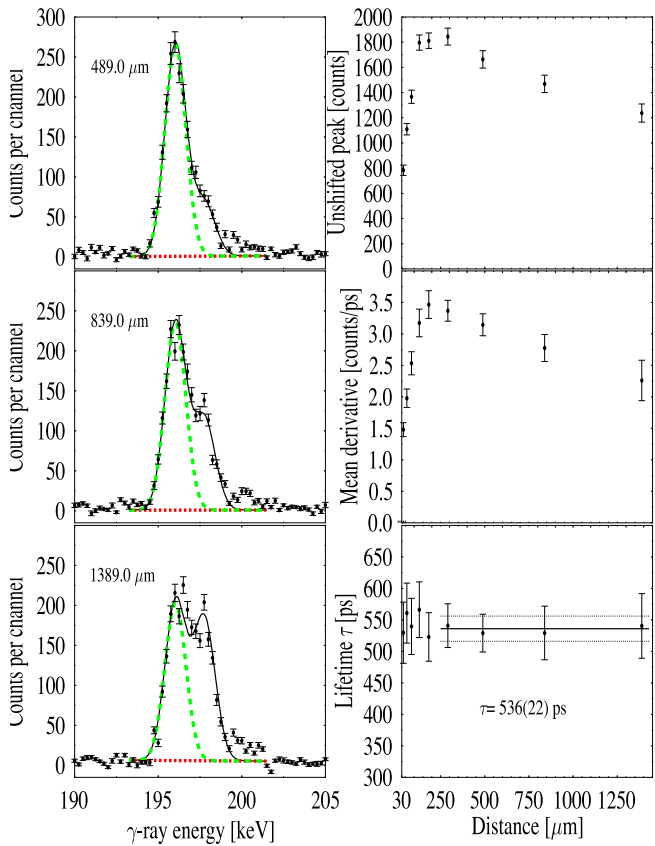}
\caption{\label{analysis_196}(Color online)  
Line-shape analysis of the {$2^+_1\to 0^+_1$,} 196 keV $\gamma$-ray  
transition and determination of the lifetime
of the $J^{\pi}$ = 2$_{1}^{+}$ state, according to the procedure presented  
in \cite{Petkov-99-A}.
On the left-hand side, spectra measured by the detectors of ring 1  
(34.6$^{\circ}$ with respect to the beam) in coincidence with a part of the 
shifted 
component of the 4$_{1}^{+}$ $\rightarrow$ 2$_{1}^{+}$ transition, also taken 
on ring 1, and their  fit (full line) are shown at three distances.
The fits of the unshifted peak and of the background are  shown with 
long-dashed line. The fitted DSA contribution ( negligible in this case) 
 is represented with a short-dashed line.
The upper panel on the right-hand side (rhs) shows the intensities of the 
unshifted component of the 196 keV transition, while the curve in the middle
 panel on  the rhs corresponds to the derivative in the  denominator of 
Eq.~\ref{normal-DDCM-tau}. 
 The points reported in the lower panel correspond to the mean life values 
deduced at the different distances.  The value of  $\tau$ corresponding to 
the best fit of results from ring 1 is shown by the central horizontal line
and the uncertainty limits are drawn through the sensitivity  region. }
\end{figure}

\begin{figure}
\includegraphics[bb=5 30 585 800,clip,width=8.64cm]
{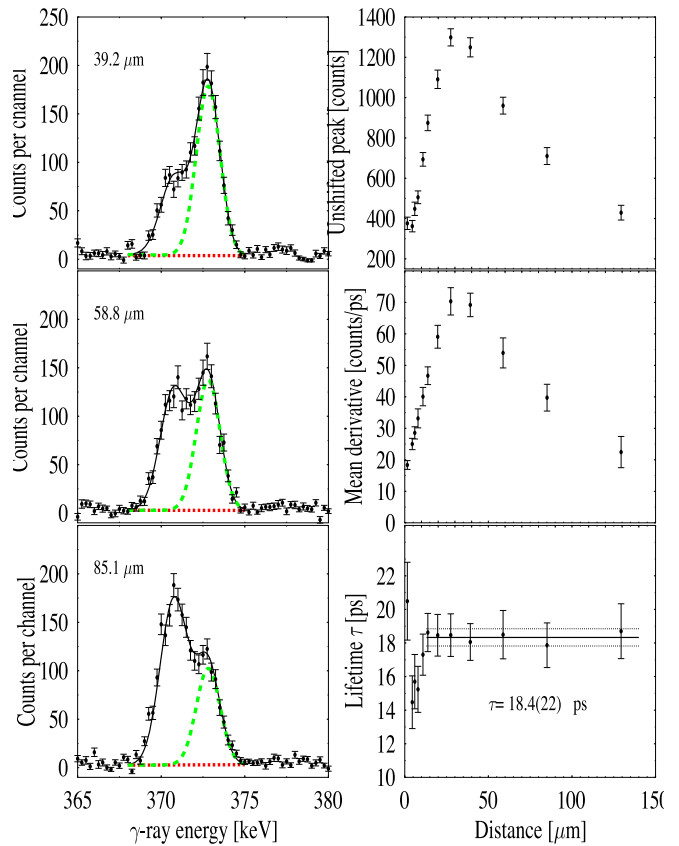}
\caption{\label{analysis_373} (Color online)  Example of the line-shape 
analysis of the $4^+_1\to 2^+_1$, 373 keV transition
 and determination of the lifetime of the $J^{\pi}$ = 4$_{1}^{+}$ state.
On the left-hand side, spectra in coincidence with a part of the shifted  
component of the 6$_{1}^{+}$ $\rightarrow$ 4$_{1}^{+}$ transition,
measured by the detectors of ring 7 (145.4$^{\circ}$ with respect to the 
beam) and their fit (full line) are shown at three distances.
The meaning of the other panels is the same as in Fig.~\ref{analysis_196}. 
 }
\end{figure}

\begin{figure}
\includegraphics[bb=5 30 585 800,clip,width=8.64cm]
{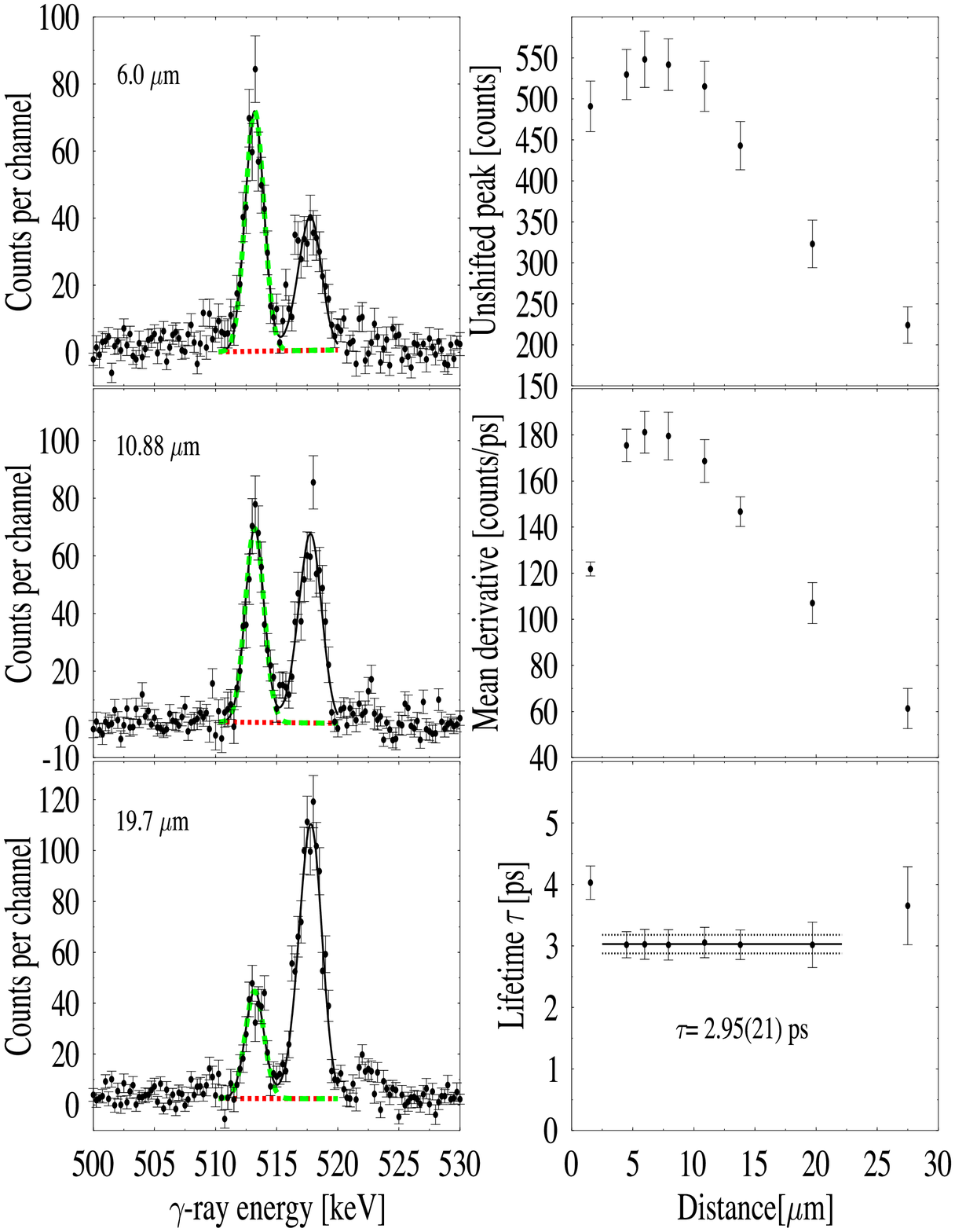}
\caption{\label{analysis_513} (Color online) 
Line-shape analysis of the $6^+_1\to 4^+_1$, 513 keV $\gamma$-ray transition
measured at three different
distances and registered by detectors from the forward ring 1 and 
determination of the lifetime of the $J^\pi=6_{1}^{+}$ state in $^{122}$Ba.
On the left-hand side the spectra obtained by setting a gate on the shifted
component of the 620 keV transition, which feeds directly the level of 
interest are shown. The meaning of the other panels is the same as in 
Fig.~\ref{analysis_196}.}
\end{figure}

\begin{figure}
\includegraphics[bb=5 30 585 800,clip,width=8.64cm]
{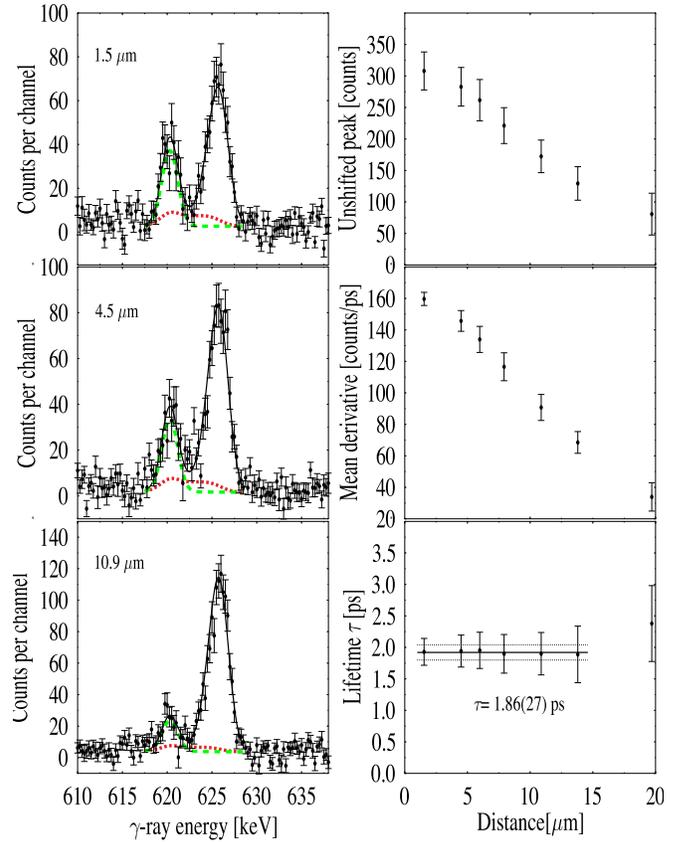}
\caption{\label{analysis_620} (Color online)  
Line-shape analysis of the $8^+_1\to 6^+_1$, 620 keV $\gamma$-ray transition
measured at three different
distances and registered by detectors from the forward ring 1 and  
determination of the lifetime of the $J^\pi=8_{1}^{+}$ state in $^{122}$Ba.
On the
left-hand side the spectra obtained by setting a gate on  the shifted 
component of the 694 keV  transition, which feeds directly the level of 
interest are shown. The importance of the  DSA contribution (short-dashed 
line) is evident.
The meaning of the other panels is the same as in Fig.~\ref{analysis_196}. 
}
 \end{figure}

Possible contaminations of the relevant transitions have been accurately 
evaluated in the analysis.
For the 196 keV, $2^+_1 \to 0^+$ transition they  could be due to the 199 
keV-band(4) 
transition or to the known 196 keV transition in $^{121}$Xe,
populated in the same reaction.
 To avoid 
overlapping of the 196 keV line, in coincidence with the shifted part of the
373 keV line, with the 199 keV-band(4)   line (which is totally shifted in 
this case) it has been sufficient to use only the coincidence spectra at the
 most forward ring. As to  the 196 keV transition from $^{121}$Xe, we have 
verified in our triple-coincidence spectra (with double gates on the 196 kev
transition and with any one of the transitions of 373 keV, 513 keV and 620 
keV) that it is not in coincidence with any transition of energy close to 
the ones used in our gates  for  RDDS analysis.

The lifetime of the $2^+_1$ level deduced from the  measurements perfomed 
at LNL is $536\pm 22$~ps, where the  errors are the statistical ones.  
With the inclusion of systematic uncertainties  we obtain  
$\tau(2^+_{1})=536\pm30$~ps. The value deduced from the Cologne data is 
fully compatible with such a result.
We observe that  the present  value is appreciably larger than that
 ($428\pm 39$ps) reported in \cite{Morikawa-92}.

An independent  evaluation  the lifetime of the  $2^+_1$ state has been 
performed by exploiting  the Integral Decay Curve Method. The data 
collected at  LNL have been analyzed  according to the formula~\cite{IDCM}
\begin{eqnarray}
\tau=\frac{ \int_{x_1}^{x_2}\{B_{s},A_{u}\}\ {\rm dx}}{
\bar{v}(x_2)\{B_{s},A_{s}\}(x_2) - \bar{v}(x_1)\{B_{s},A_{s}\}(x_1)}
\end{eqnarray}
extended to the full range of distances.  The result ($530\pm 30$ ps)
   is in excellent agreement with the one obtained with the DDCM method.

In the analysis concerning the lifetime of the  $4^+_1$ state we have taken
into account  the possible bias for the $4^+_1\to 2^+_1$ transition, due to
the fact that the gate on the shifted part of the 513 keV line includes also
the shifted part of the 514 keV-band(4) transition. As only the most shifted
part of the 513 keV line has been included in the gate, the fraction of the
shifted 514 keV-band(4) line present in this gate can even be somewhat larger
than 5\%. The largest bias would correspond to a situation in which the 514
 keV-band(4)  line is fully Doppler shifted while the daughter transitions 
feeding the $4^+_1$ level are completely unshifted. In this extreme situation,
the apparent mean life of the $4^+_1$ level could be from 10\% to 20\% longer
than the true one, depending on the distance. We therefore evaluated as 20\%
 the error of $\tau (4^+_1)$, on the side of lower values.

Particular attention has been payed to the evaluation of the lifetime of the
$6^+_1$ state which, as observed in the previous section, can
be affected by the presence of the 514 keV-band(4)  transition.  In 
coincidence with the  {\em shifted part } of the 620 keV line, only the
{\em shifted part} of the 514 keV-band(4) line is present. It overlaps with
the shifted part of the 513 keV line, whose unshifted part remains unaltered.
Therefore, the {\em apparent} time derivative of the shifted intensity of the
513 keV line will include, in addition, the corresponding derivative of the
shifted 514 keV-band(4) line. From the value $1.6\ 10^{-2}$ of the ratio of
the 514 keV-band(4) to the 513 keV $\gamma $ ray, in coincidence with the 620
 keV transition, deduced in the previous section,  the estimated maximum
possible bias on the resulting lifetime turns out to be less than about 2\%,
much less than the estimated limit of systematic uncertainty. As seen in 
Fig.~\ref{F:3}, there is a small contamination due to the presence of  the
511 keV peak. This fact was also the reason for choosing to analyze only 
the two forward rings, where  the shifted peaks are at higher energies.

 As to  lifetime of the $8^{+}_{1}$ state, an inspection of
 Fig.~\ref{analysis_620} reveals  the importance of taking into account the
 DSA contribution  for a precise analysis of the data for short lifetimes.

The attempt to determine the mean life of the $10^+_{1}$ level was based on
 the analysis of a few results at the shortest distances, which are very 
sensitive to the evaluation of the background. As a consequence, we can only
give an upper limit to the mean life of this level. The results relative to
the DSA analysis of the data concerning the $10^+_1$ state, which set a lower
 limit  on the lifetime of this level, will be presented  in a forthcoming 
paper\cite{mic2010}.

  The values of the  lifetimes deduced in the present work are shown in
 Table~\ref{table_values} together  with the upper limit deduced for  the
  $10^+_{1}$ level.  Only for the  2$^{+}_{1}$ state a previous  value
 ($428\pm 39$ps) was known \cite{Morikawa-92}. 

 The statistical errors on the  lifetimes are less than 5\% for all the 
 states of interest.   
The error limits reported in the table include a conservative estimate of the 
systematic uncertainties, mostly related to  the uncertainty  in the 
stopping power. The error limits reported in the tables include an
estimate of additional uncertainties related to the background subtraction.

\begin{table}
\caption{\label{table_values}
Values of the  lifetimes of excited levels of $^{122}$Ba determined  in the
present work, without ($\tau_{nc}$) and with ($\tau_{corr}$) the correction
for the finite slowing-down time of the recoils in the stopper. For the
 6$^{+}_{1}$, the reported value is the average
of results obtained from rings 1 and 2; for the 4$^{+}_{1}$ and 8$^{+}_{1}$
 states  the average is  taken over rings 1, 2, 6 and 7.  Spin and  parity 
of the relevant states are shown in column 1. Level and transition energies
are presented in column 2 and 3, respectively. }
\begin{ruledtabular}
\begin{tabular}{cccccc}
& State & Energy & E$_\gamma$& $\tau_{nc}$ & $\tau_{corr} $\\
& $J^{\pi}$ & [keV] & [keV] & [ps] & [ps] \\[0.5mm]
\hline
& 2$_{1}^{+}$ & 196   & 196      & 536$\pm$ 30  &  \bf 536$\pm$ 30  \\
& 4$_{1}^{+}$ & 569   & 373      & 18.4$^{+2.2}_{-3.7}$ & 
\bf 18.4$^{\bf +2.2}_{\bf -3.7}$ \\
& 6$_{1}^{+}$ & 1082  & 513      & 2.95$\pm$0.21 & \bf 2.95$\pm$0.21 \\
& 8$_{1}^{+}$ & 1702  & 620      & 1.60$\pm$0.12 & \bf 1.86$\pm$0.27\\
& 10$_{1}^{+}$ & 2396 & 694      & $~$ & \bf $\boldmath{<}$ 1.62 \\
\end{tabular}
\end{ruledtabular}
\end{table}

\section{Discussion}

At present, one of the most important  topics  in the study of the nuclear 
structure concerns  the nuclear shape/phase transitions.
Its investigation can be performed  in the framework of both the collective 
model \cite{boh75} and the interacting boson approximation  (IBA) 
model \cite{iach87}.  In the latter  case a transition  between 
U(5) and SU(3)  symmetries corresponds  to a change  from a spherical to 
an axially deformed shape. 
The evolution of observables in the transitional  region exhibits a rather 
discontinuous  behavior and in general its detailed  study requires a 
numerical solution of the  Hamiltonian.
However,  the X(5) model proposed by Iachello \cite{iac02}, based on a 
simplified Bohr Hamiltonian (in which the $\beta$ and $\gamma$ degrees of 
freedom are decoupled  and are associated to  a square  well potential  and
 a harmonic oscillator potential,  respectively) provides an analytic 
solution which simply describes nuclei near the critical point of the 
spherical to axially-symmetric rotor transition. 

The first identification of nuclei with an  X(5) character was that of 
$^{150}$Nd \cite{Kruecken-02} and $^{152}$Sm \cite{Casten-01}, both having
 N = 90. The analysis has  been soon after  extended to  other nuclei around 
 A $\approx$  150 and  to other mass regions. It has been found that, in many
 cases, nuclei having the excitation energies of the g.s. band  in  good 
agreement with the predictions of the X(5) model  do not have  B(E2) values 
 displaying the  trend predicted by the model.  In particular, in some cases,
 the B(E2) ratios  are close to those of the SU(3) limit  (see, 
e.g. \cite{clark03,dew04,mccu05,mcc06} and references therein).

A possible identification of $^{122}$Ba as a close-to-X(5) nucleus   has 
been proposed by Fransen {\it et  al.}~\cite{Fransen-04}. Indeed,  
   the energy ratio R$_{4/2}$ = $E(4_1^+) / E(2_1^+)$  in the N = 66  
isotones  (see Fig.1) takes in this nucleus the value  2.9 predicted by 
the model. Moreover,  the  relative excitation  energies of  the ground state
 band as a function of J  are in very  good agreement with the  X(5) 
predictions. 
  In \cite{Fransen-04} a tentative identification of the { whole $K$ = 2 
$\gamma$-band has been reported 
  and  its decay properties have been compared  
   with the predictions of the X(5) model,  the O(6) symmetry, 
 and the rotor model. 
 The assignment of the  even-J levels  of the proposed $\gamma$-band to  
$^{121}$Xe, made in the present work,  limits the effectiveness of this 
comparison. }
Anyway,  no definite conclusion about a possible X(5) character of 
 $^{122}$Ba was drawn in \cite{Fransen-04} and  the final remark   was that 
 this nucleus might have a structure  more complicated than  expected.  

The results obtained in the  present work allow to extend the  analysis to 
the electromagnetic  transitions probabilities, which provide a very critical
 and necessary test  to establish the structure of  a nucleus. 

In addition to the X(5) model we have considered the IBA model \cite{iach87} 
in both the IBA-1 version, where no distinction is made between proton and
 neutron bosons,  and the  IBA-2 version, where they are considered separately.

To investigate the  shape/phase transitions in the framework of the IBA-1
 model we  used the two-parameter Hamiltonian \cite{zam02,wer02} 

\begin{equation} 
 H =  c [(1-\zeta) \, (\hat n_{d})+ (\zeta/4 N_{B}) \, \hat  Q^{(\chi)} 
\cdot Q^{(\chi)} ]
\label{E:3}
\end{equation}
where c is a normalization factor, $\hat n_{d} = d^{\dag} \cdot \tilde{d}$, 
 $ Q^{(\chi)} = (s^{\dag}  \tilde{d} + d^{\dag}  \tilde{s})$ + $\chi
 (d^{\dag} \tilde{d})$ and N$_{B}$ is the number of the valence bosons.
 The same parametrization of the boson quadrupole operator is used in both
 the Hamiltonian and in the E2 operator (Consistent Q 
Formalism \cite{war83,lip85}).
 The U(5) limit is obtained for $\zeta$= 0 while  the SU(3) and O(6) limits
 are obtained for   $\zeta$ = 1 and for $\chi   = -\sqrt{7}/2$ and 0,
 respectively. 

The calculations have been performed  following  the same line
 of  \cite{mcc05} and for  a boson number  N$_{B}$ = 11, corresponding to
 that of $^{122}$Ba,  which has  6 protons and 16 neutrons valence nucleons.
  We first varied  $\zeta$ from 0  to 1,  giving  the parameter  $\chi$  
three different values:   $-1.32 ~ (= -\sqrt{7}/2$), $-0.75$, and $ -0.20$.
 To display the  evolution of the nuclear structure for  the three cases  we
  report  in  Fig. 10 the values of $R_{4/2}$  (which is one of  the usual
 benchmarks adopted to characterize the degree of deformation of a nucleus) 
 as a function of $\zeta$.  In the figure, the vertical lines mark the values
 of  $\zeta$ (0.562, 0.65, and 0.92) where the three curves attain the  value
  predicted by  the X(5) model.

By referring to the so-called Casten triangle \cite{cas81}, reported in the
 inset of Fig. 10, the $\chi  =  -1.32$ transition is along the U(5) - SU(3)
 leg.  For $\chi = -0.75$, and $-0.20$ the trajectories are inside the 
triangle, along  lines whose polar coordinates can be deduced from the 
expressions given in \cite{mcc04}, which convert the parameters $\zeta$ and 
 $\chi$ into the parameters $\rho$ and $\theta$. These polar coordinates 
allow to span the entire  triangle with $\rho$ ranging from 0  [U(5) vertex]
  to 1 [SU(3) and O(6) vertices] and $\theta$  from 0$^{\circ}$ [SU(3) vertex]
 to 60$^{\circ}$  [O(6) vertex].    
The  values of the parameters $\rho$ and $\theta$, which correspond to the 
 values of  $\chi$ and $\zeta$  $(-1.32, 0.562), (-0.75, 0.65)$, and $(-0.20,
 0.92)$  reproducing the X(5) prediction for 
$R_{4/2}$, are (0.56, 0$^{\circ}$),     (0.56, 26$^{\circ}$), and (0.85,
 51$^{\circ}$), respectively.   The latter case corresponds to a  quite 
 $\gamma$-soft nucleus.  

 \begin{figure}[h]
\hspace*{-2mm}
\includegraphics[width=86mm]
{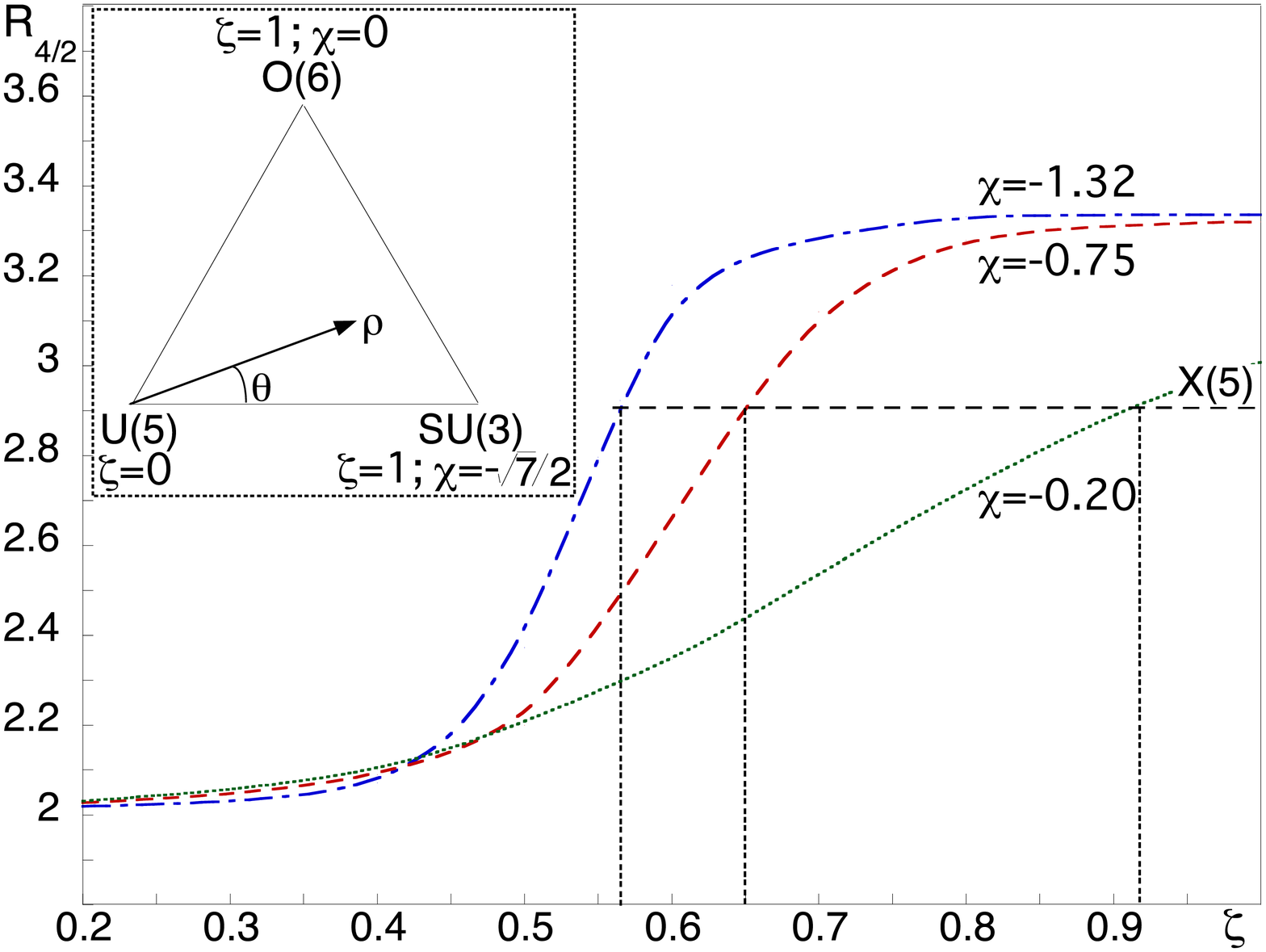}
\caption{\label{F:10} (Color online) 
 The ratio $R_{4/2}$ is shown  as a function of  $\zeta$ for the three 
indicated values of  $\chi$ and   N$_{B}$ = 11. The horizontal line shows
 the value predicted by the X(5) model, the vertical lines mark the values 
of $\zeta$ for which $R_{4/2}$ takes the value predicted by the X(5) model.
 In the inset the Casten triangle is displayed.}
\end{figure}  
   
As a second step, we calculated  the values of the energies and $B(E2)$ 
values of the g.s. band for the three pairs of values of $\chi$ and $\zeta$ 
just mentioned.
  The  $E(J)/ E(2)$ and   $B(E2; J \to \! J-2)/B(E2; 2_{1}^{+} \to 
\! 0_{1}^{+})$ ratios are shown, as a function of J, in Fig. 11, together
 with  the predictions of the X(5) model and the U(5) and SU(3) limits.  
It is seen that the IBA-1 values of E(J)/E(2), which  are very close for the
 three cases considered,  are slightly higher than those of the X(5) model  
and that the normalized $B(E2)$ values slowly decrease from [a] to [c]  and
 approach the values of the SU(3) limit. 

 \begin{figure}[h]
\hspace*{-3mm}
\includegraphics[width=86mm]
{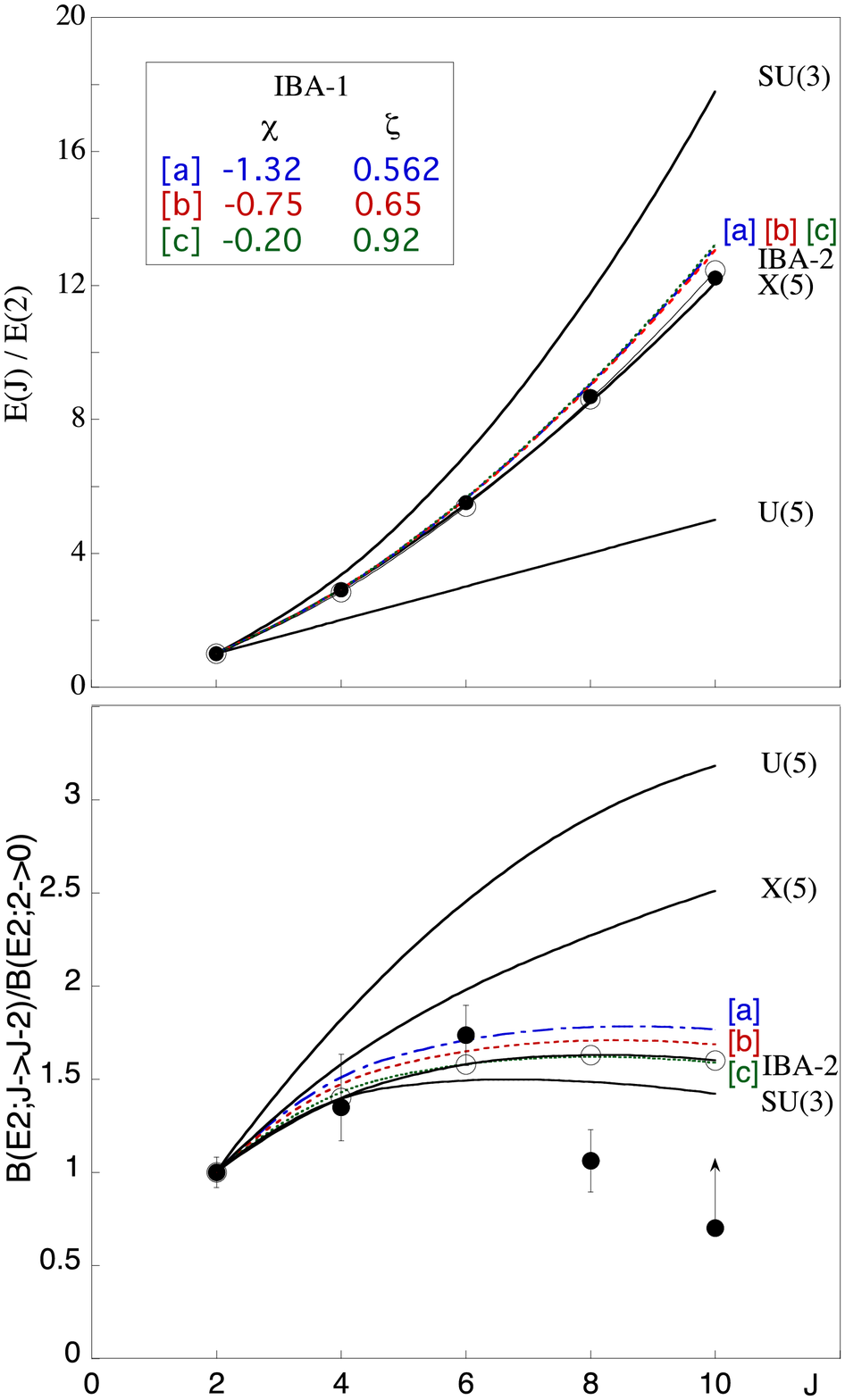}
\caption{\label{F:11} (Color online)  
 The experimental values (black full circles) of  the  energy ratios  
(upper part)   and B(E2)  ratios (lower part) of the g.s. band of $^{122}$Ba 
 are compared with the X(5), U(5) and SU(3) predictions (full lines), with the
  IBA-1 calculations (dashed lines [a], [b] and [c])
  and with the IBA-2 calculations (open 
circles).  The values of  the parameters $\chi$ and $\zeta$ used in the IBA-1
 calculations  are given in the inset.}
 \end{figure}  

As to the analysis in the framework of the IBA-2 model, we kept the 
Hamiltonian parameters fixed to those found in the studies of the whole Te,
 Xe, Ba isotopic chains  \cite{otsu93,mizu96}. In this way  a possible  
agreement   due to a fine tuning of  the model parameters to the  experimental
  data can be excluded. 
We used the same  Hamiltonian as  in ~\cite{otsu93} with the addition of  the
 Majorana term $\hat M_{\pi\nu}$
\begin{equation} 
 H =\varepsilon\,(\hat  n_{d_\pi}+\hat n_{d_\nu})+
\kappa\, \hat  Q_\pi^{(\chi_\pi)}\cdot \hat  Q_\nu^{(\chi_\nu)} + 
\hat M_{\pi\nu}(\xi_1,\xi_2,\xi_3) \label{first}
\label{E:3b}
\end{equation}
The  Majorana parameters $\xi_i$ have been kept at 1 MeV so as to push at 
high energies states non symmetric in the proton-neutron degrees of freedom. 
The values adopted for the parameters  $\varepsilon$ (0.5 MeV) , 
$\kappa$ (- 0.21  MeV), $\chi_\nu$ (0.1),  $\chi_\pi$ (- 0.5) are those 
deduced  in~\cite{otsu93} for $^{122}$Ba. 
The values of $\chi_\nu$ and $\chi_\pi$ in the E2 transition operator
\begin{eqnarray}
\label{E:4}
\hat T(E2)
&  = &
e_\nu \hat Q_\nu^{(\chi_\nu)} +  e_\pi \hat
Q_\pi^{(\chi_\pi)}\label{tedue}
\end{eqnarray}
 are the same as in the  Hamiltonian (CQF  \cite{war83,lip85}). 
The proton- and neutron-boson effective charges, for which no value is 
reported in \cite{otsu93,mizu96} for $^{122}$Ba,  have been kept at the
 values  $e_\pi=0.18$ {\it e}b and $e_\nu=0.095$ {\it e}b, respectively.
   They are  slightly higher than those evaluated  microscopically 
in~ \cite{mizu96} for  the isotone $^{120}$Xe. {   It is important to
 notice that  a   50$\%$ variation  of  the parameters $e_\pi$ and/or  
$e_\nu$  changes by less than $\sim 1\%$ the relevant  $B(E2)$ ratios so
 that,  in  this comparison, basically no free parameters were used. }

\begin{table}[h]
\caption{\label{table_values2}
Values of the reduced transition strengths $B(E2)$ in the g.s. band of 
$^{122}$Ba, compared with the results of the IBA-2 calculation described in the
text, and with the predictions of the X(5) model ( which has been normalized
 to the experimental value for the $2^+_1\rightarrow 0^+_1$ transition).}
\begin{ruledtabular}
\begin{tabular}{cccccc}
 & Transition & $E_\gamma$ &  $B(E2)${\large $_{exp}$} & $B(E2)_{IBA-2}$ & 
$B(E2)_{X(5)}$  \\
&                     & [keV]                & [{\it e}$^2$b$^2$]         
             & [{\it e}$^2$b$^2$]      & [{\it e}$^2$b$^2$]   \\
\hline 
& 2$_{1}^{+}\rightarrow 0_1^+$  & 196        & 0.45$\pm 0.03$           
               & 0.43                     &          0.45\\
& 4$_{1}^{+}\rightarrow 2_1^+$  & 373        & 0.60 $^{+0.12}_{-0.07}$  
         & 0.62                     &           0.72 \\
 & 6$_{1}^{+}\rightarrow 4_1^+$  & 513       &  0.77$\pm 0.06$           
              & 0.69                     &           0.89\\ 
& 8$_{1}^{+}\rightarrow 6_1^+$  & 620       &  0.47$\pm  0.07$           
              & 0.71                     &           1.03\\ 
&10$_{1}^{+}\rightarrow 8_1^+$ & 694       &    $\ge $  0.46             
                  & 0.70                     &            1.13\\
\end{tabular}
\end{ruledtabular} 
\end{table}

The $B(E2)$  values calculated in the framework of  the IBA-2 model and  
those predicted by the X(5) model are compared to the experimental ones  
in {Table II.  

 The  energies }ratios and the  $B(E2)$ ratios  are shown   in Fig. 11 as a 
 function of J. 
It is seen that the  agreement between experimental and calculated  energy 
ratios is comparable to   that obtained by the X(5) model. The trend of the 
calculated $B(E2)$ ratios is similar to that obtained in the  IBA-1 
calculations (with  values   very close  to those corresponding to parameters
 [c] in IBA-1) and that of  both  IBA  predictions is close to that of the 
SU(3) model. The IBA calculations are thus able to describe  the energy ratios
 at a level close to   that of the X(5) model   and the  $B(E2)$ ratios at a 
level not too far from that  of the SU(3) symmetry, which, however, 
overestimates the $B(E2)$ ratio of the $8_{1}^{+}$  state.  On the other hand,
 the predictions of the X(5) model, which are   compatible with the 
experimental B(E2) ratios up to J = 6, are hardly compatible  with the 
experimental value  for J = 8.

  A precise value of the lifetime of the $10^{+}_{1}$ state would be 
necessary to clarify whether (and how much) the trend of  $B(E2)$ ratios  
keep decreasing at   higher spin values. This, indeed, would allow to perform 
a more stringent comparison of experimental and theoretical data.

\section{Conclusions}

The lifetimes of the $4^+_1,\ 6^+_1,\ 8^+_1$  levels and  an upper limit for 
the lifetime of the $10^+_1$ level of 
$^{122}$Ba have been  determined for the first time in the present work. As to
 the $2^+_1$ state, the present mean life 
$\tau_m =536\pm 30$~ps is appreciably longer than that  
($\tau_m =428\pm 39$~ps) obtained
by Morikawa {\it {\it et al.}}~\cite{Morikawa-92} by electronic timing.

It has been found that the 587-422-196 keV cascade, tentatively interpreted 
in~\cite{Fransen-04} as the  $4_2^+\rightarrow 2_2^+\rightarrow 2_1^+$ 
cascade in $^{122}$Ba, belongs to the $^{121}$Xe  nucleus. As to the 
$^{122}$Ba, two more levels have been added on top of the band based on the 
 $3^{+}_{1}$ state.

A description of $^{122}$Ba as an X(5)-like nucleus matches  very well the 
behavior  of the experimental energies, whereas the  data on the B(E2) ratios
  of the g.s. band obtained in the present work suggest a picture closer to 
the SU(3) description (which in turn is  not compatible  with the data on the 
energies).   Both IBA-1 and IBA-2 calculations  predict values comparable to
 those of X(5) for the energies ratios and close  to the SU(3) for the $B(E2)$
 ratios.  To test  to what extent a description of  the 
$^{122}$Ba nucleus in the framework of the IBA model is close to the 
experimental data a precise experimental value  of the lifetime  of the 
$10^{+}_{1}$ state would be very valuable. 

 \section{acknowledgments}

This research has been supported within the ``Research Infrastructure
Action under FP6 - Structuring the European Reaserach Area - Programme, 
Contract n. 506065
EURONS - EUROpean Nuclear Structure research''. D.T. and P.P. are indebted to
the National Science Fund at the Bulgarian Ministry of Education and
Science under contract number RIC-02/2007 for a financial support.

\end{document}